# Long-term variation of population exposure to PM2.5 in Eastern China: A perspective from SDG 11.6.2

Yuheng Zhang [1], Qi Zhou [1], Ke Chang [1*]

1   School of Geography and Information Engineering, China University of Geosciences,

Wuhan, P.R.China

email: hengxin@cug.edu.cn; zhouqi@cug.edu.cn; changke@cug.edu.cn

**Abstracts:** Air pollution (e.g., PM2.5) has a negative effect on human health. Recently, the population-weighted annual mean PM2.5 concentration ($PWAM$) has been selected as an indicator 11.6.2 in Sustainable Development Goals (SDGs), for various countries to perfrom a long-term monitoring of population exposure to PM2.5 in cities. However, few studies have employed this indicator for a city-level analysis and also in a long-time series (e.g., for decades). To fill this research gap, this study investigates the long-term (2000-2020) variation of population exposure to PM2.5 in Eastern China (including 318 prefecture-level cities). Three categories of open geospatial data (including high-resolution and long-term PM2.5 and population data, and administrative boundary data of cities) are involved for analysis. We found that: 1) A considerable decrease has been observed for the $PWAM$ during 2014-2020. 2) In 2020, the $PWAM$ is for the first time lower than the interim target-1 (35 μg/m$^3$) defined by the World Health Organization for 214 prefecture-level cities in Eastern China, which accounts for 67% of the total population. The results indicates a considerable improvement of air quality in Eastern China. More important, this study illustrates the feasibility of using open geospatial data to monitor the SDG indicator 11.6.2.

# 1. Introduction

Air pollution has a significantly negative effect on human health. It has been estimated that in 2015 more than 4.2 million premature deaths worldwide were attributed to air pollution (Cohen et al. 2017), especially due to exposure to fine particulate matter with a diameter less than 2.5 microns (PM2.5). The level of population exposure to PM2.5 has also been selected by the United Nations as an indicator in Sustainable Development Goals (SDGs), i.e., SDG 11.6.2: "Annual mean levels of fine particulate matter (e.g., PM 2.5 and PM 10) in cities (population weighted)", in order to achieve the Goal 11: "Make cities and human settlements inclusive, safe, resilient and sustainable" (Klopp and Petretta 2017; United Nations 2020). It is therefore desirable for various countries to perform a long-term monitoring of this indicator at not only national but also city levels.

Extensive studies have paid attention to the analysis of PM2.5 concentration for a long time span (e.g., over decades) (Zhao et al. 2019; Lim et al. 2020; Wei et al. 2021). For instance, Zhao et al. (2019) analyzed the temporal-spatial variation of PM2.5 concentration in China during the period 1999-2016. However, these studies did not take the distribution of population into consideration. Some studies have assessed the level of population exposure to PM 2.5 (Yao and Lu 2014; Zhang and Cao 2015; Guo et al. 2017; Cohen et al. 2017; Chen et al. 2018). But, most of these assessments were carried out at a national or regional level rather than at a city level; on the other hand, most of them only involved a short time span (e.g., a calendar year) for analysis. To be best of our knowledge, few studies have employed the SDG indicator 11.6.2 to perform a city-level analysis and also for a long time span.

To fill this research gap, this study aims to investigate the long-term (2000-2020) variation of population exposure to PM 2.5 in Eastern China (including 318 prefecture-level cities). We selected China as the study area because the number of PM2.5-realted premature deaths in this country was estimated to 1.1 million in 2015, which accounted for 26% of the number (4.2 million) in the world (Cohen et al. 2017). Therefore, a long-term monitoring of population exposure to PM2.5 in China has received much attention worldwide (Guo et al. 2017; Zhang et al. 2019; Zhao et al. 2019; Wei et al. 2021). More important, both high-resolution and long-term PM2.5 and population data products have recently been published for conducting this study.

## 2. Study area and Data

The study area includes 318 prefecture-level cities in Eastern China (Figure 1).

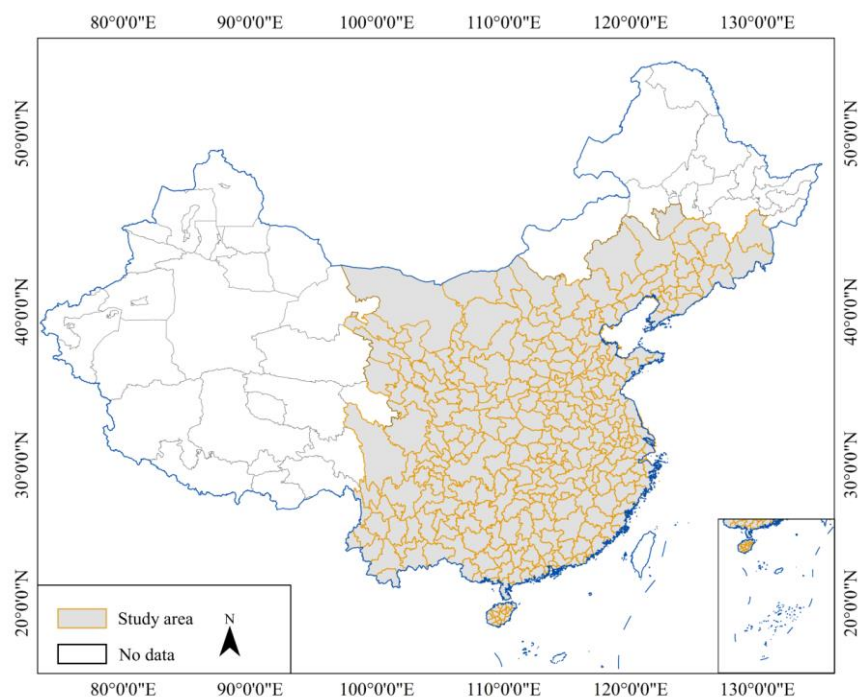

Figure 1. Study area.

Specifically, this study area has a population of 1.36 billion, which accounts for 94.8% of the total population in China. Moreover, both high-resolution and long-term PM2.5 and population data are freely acquirable for this study area.

Three categories of data sources are involved for the analysis.

1) PM2.5 data: A high-resolution and long-term yearly PM2.5 data product was recently made for public use[1] (Wei et al. 2021). The data product covers the region of Eastern China. Also, it has a spatial resolution of 1km, and it includes a long time series of datasets during 2000-2020.

2) Population data: The global 100m-resolution population data product (Worldpop[2]) was acquired (Lloyd et al. 2019; Zhang et al. 2022). The data product also includes a long time series of datasets during 2000-2020. Moreover, it has been used for the evaluation of SDG 9.1 (Li et al. 2022) and SDG 11.7 (Long et al. 2022; Zhou et al. 2022). Thus it is possible to investigate the variation of population exposure to PM2.5 not only for decades (21 years) but also in the recent years.

3) City boundary data: The administrative boundary data of 318 prefecture-level cities in Eastern China were freely acquired from the National Catalogue Service For Geographic Information[3].

## 3. Methods

The SDG indicator 11.6.2 denotes the population-weighted annual mean PM2.5 concentration ($PWAM$). This indicator is calculated as follows:

$$PWAM = \frac{\sum_{i=1}^{k} p_i \times c_i}{\sum_{i=1}^{k} p_i}$$

where, $c_i$ denotes the PM2.5 concentration in the 1km grid cell $i$. $p_i$ denotes the total population in the same grid cell $i$, i.e., it equals to the total population in 100m-resolution population grids whose centroids are located inside the grid cell $i$. $k$ denotes the number of 1km grid cells in a geographical region. The $PWAM$ can be calculated at a regional level (e.g., the whole study area) and also at a city level (e.g., each prefecture-level city).

The specific experiment steps include: First, the $PWAM$ was calculated and plotted for the whole study area during 2000-2020. Second, the $PWAM$ for each of the 318 prefecture-level cities was also calculated and mapped for these years. Third, all the prefecture-level cities were divided into five different intervals (i.e., 0-35; 35-50; 50-75; 75-100; >100 μg/m³), and the population percentages of various intervals were calculated and plotted for the whole study area during 2000-2020.

## 4. Results and analyses

Figure 2 plots the variation (2000-2020) of the $PWAM$ for the whole study area. This figure shows that: the variation has approximately passed through three phases: In the first phase (2000-2003), the $PWAM$ increases from around 60 μg/m³ (in 2000) to 70 μg/m³ (in 2003). In the second phase (2004-2013), the $PWAM$ fluctuated slightly around 70 μg/m³, and it reaches to the maximum (72 μg/m³) in 2011. In the third phase (2014-2020): the $PWAM$ decreases year by year, and it decreases to 34 µg/m³ in 2020. It is worth pointing out that in 2020, the $PWAM$ is for the first time lower than the interim target-1 (35 µg/m³) defined by the WHO (World Health Organization 2006).

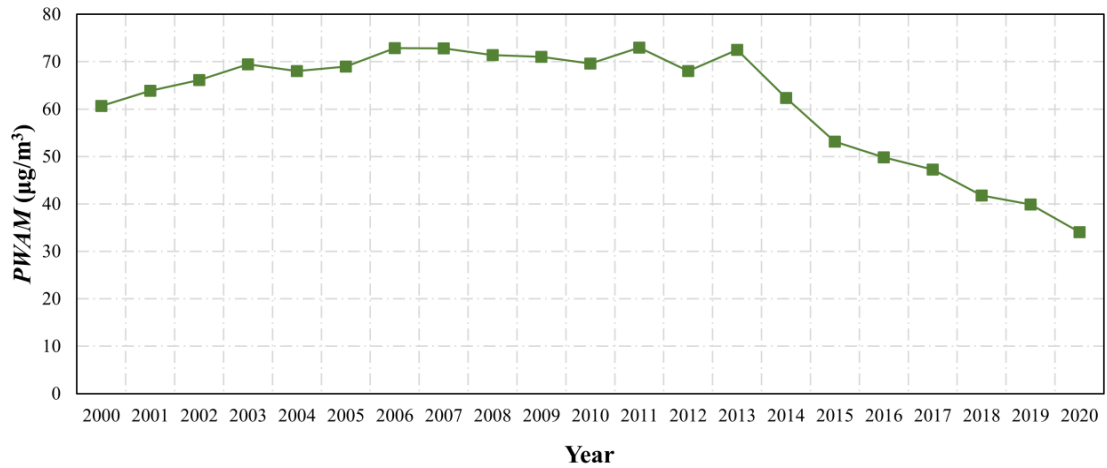

Figure 2. The variation (2000-2020) of the $PWAM$ for the whole study area.

Figure 3 shows the temporal-spatial variations (2001-2020) of the 318 prefecture-level cities. This figure shows that: In 2001, the $PWAM$ is higher than 75 µg/m³ for 45 out of the 318 prefecture-level cities. The number of such cities increases to the maximum (100) in 2011. But in 2020, there is no prefecture-level city whose $PWAM$ is higher than 75 µg/m³. On the contrary, in 2020, the $PWAM$ is lower than the interim target-1 for 214 of the 318 prefecture-level cities. The above results indicates a considerable decrease of the $PWAM$ for most of the prefecture-level cities.

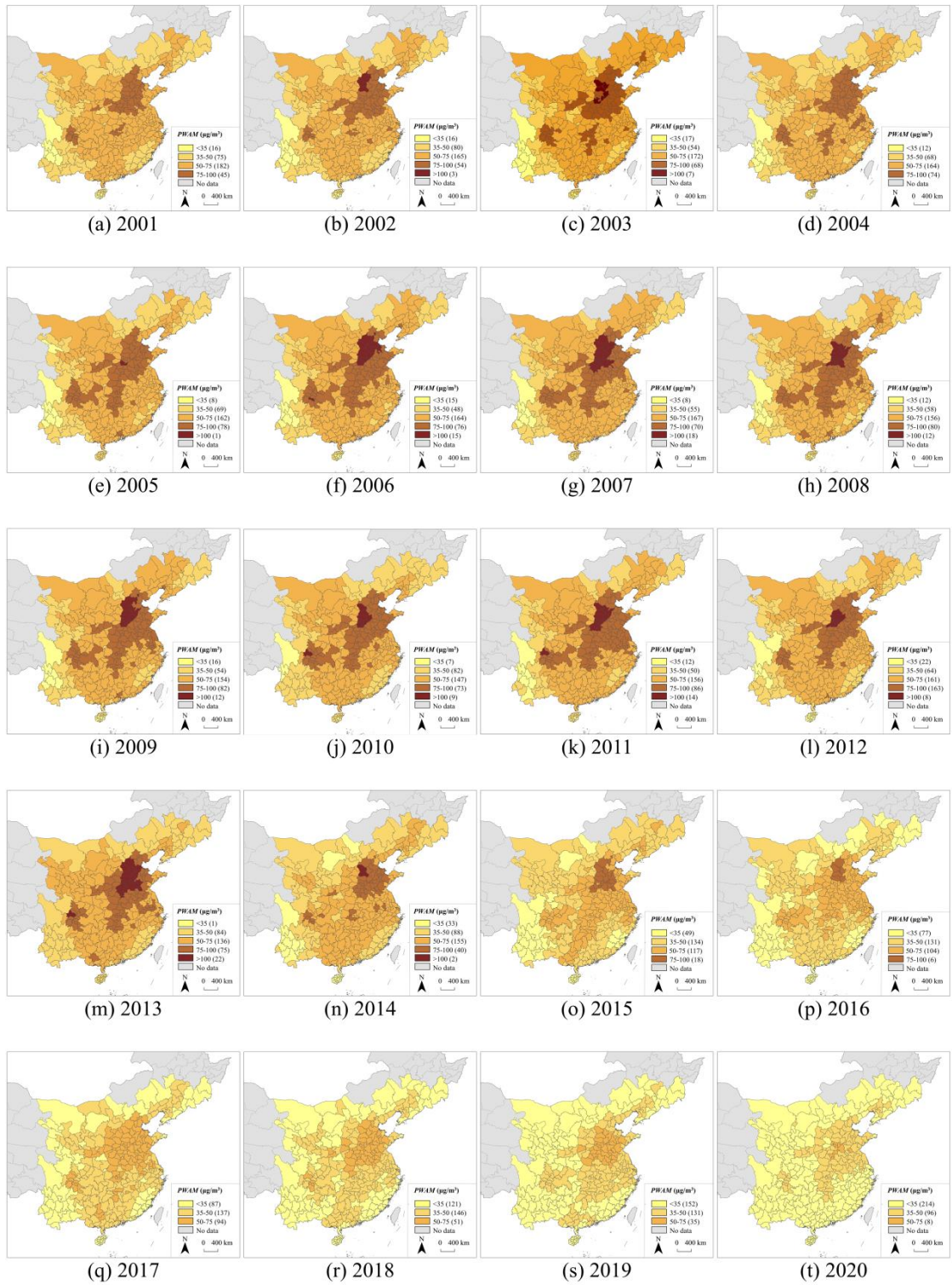

Figure 3. Temporal-spatial variation (2001-2020) of the $PWAM$ for 318 prefecture-level cities in Eastern China.

Figure 4 plots the population percentages of various PM2.5 concentration intervals for the whole study area during 2000-2020. In 2000, 65% of the total population are lived in the prefecture-level cities whose $PWAM$ is higher than 50 μg/m³. Such a percentage has reached to the maximum (88%) in 2011. However, in 2020, this percentage has been decreased to 3%. On the contrary, less than 3% of the total population are lived in the prefecture-level cities whose $PWAM$ is lower than 35 μg/m³ during 2000-2013. But in 2020, the percentage has been increased to 67%. This indicates a considerable improvement of population exposure to PM2.5 in Eastern China.

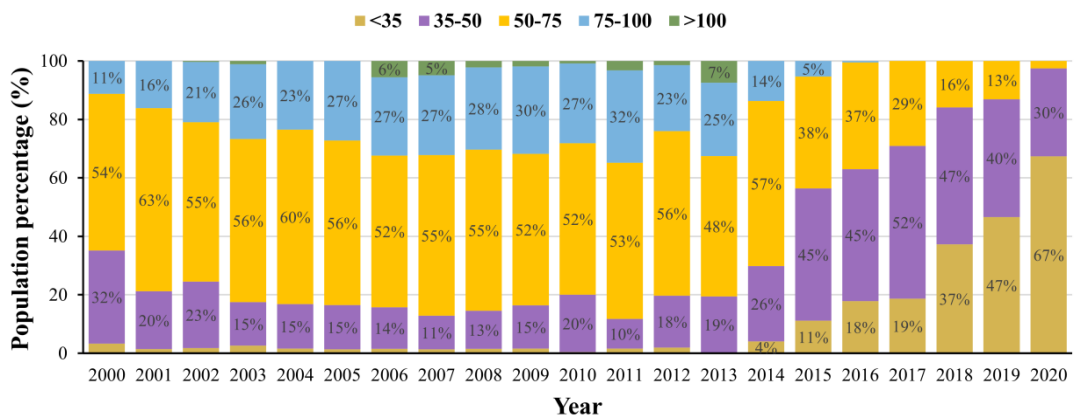

Figure 4. The variations (2000-2020) of population percentages within various PM2.5 concentration intervals (for the whole study area).

## 5. Lessons learned from this study

*5.1 Monitoring SDG indicator 11.6.2 with open geospatial data*

This study involved three categories of open geospatial data for monitoring the SDG indicator 11.6.2.

First, a high-resolution (1km) and long-term (2000-2020) PM 2.5 data product was acquired for analysis (Wei et al. 2021), but this product was only available in Eastern China. As an alternative, it may also be possible to use the global annual PM2.5 grids[4], which not only has a global coverage but also is characterized by a fine resolution of 0.01 degrees (i.e., approximately 1.1km at the equator). Nevertheless, the global data product only includes datasets during 1998-2016. Those (datasets) of recent years (e.g., 2017-2020) may not be available.

Second, the 100m-resolution global population data product (WorldPop) was used because it includes datasets of the same period (2000-2020). It is also possible to use other global population data products. As an example, the LandScan is such a data product[5], which also includes datasets for decades (2000-2019), although it has a relatively low spatial resolution (i.e., approximately 1km at the equator).

Third, the administrative boundary data of various prefecture-level cities were produced by the National Catalogue Service For Geographic Information. As an alternative, it is possible to use the Urban Centre Database GHS-UCDB[6], which includes more than 10,000 urban centres worldwide.

In summary, various categories of open geospatial data have become more and more available. This study illustrates the feasibility of using open geospatial data to monitor the SDG indicator 11.6.2. Different dataset(s) may be applied to perform a long-term analysis of population exposure to PM2.5 in other countries and regions.

*5.2 A considerable improvement of population exposure to PM2.5 in Eastern China*

A considerable improvement of population exposure to PM2.5 has been observed in Eastern China. Specifically, for the first time, the $PWAM$ has reached to the interim target-1 in 2020, not only for the whole study area, but also for its 214 prefecture-level cities. The number (214) of such cities is almost ten times than that (i.e., no more than 22) during 2000-2013. This improvement may be attributed to the Air Pollution Prevention and Control Action Plan of China, released in 2013 (Cai et al. 2017). Nevertheless, the $PWAM$ is higher than the interim target-1 in 2020, for 104 prefecture-level cities, which are mostly located in North China. Thus it is still needed for local government(s) to implement tougher policies to reduce air pollution. On the other hand, the decrease of the $PWAM$ in 2020 may somehow be attributed to the lockdown of Chinese cities after the outbreak of COVID-19 (Chauhan and Singh 2020). Thus it is still needed to perform a long-term monitoring of the SDG indicator 11.6.2 in future, in order to investigate the effect of air quality improvement. Besides, it is also interesting to investigate the significant driver(s) behind the air quality improvement during 2014-2020, which may be beneficial for other countries and regions to reduce air pollution.

## 6. Conclusion

This study conducted a case study of using the SDG indicator 11.6.2 to perform a long-term (2000-2020) analysis of population exposure to PM2.5 in Eastern China. Specifically, the population-weighted annual mean PM2.5 concentration ($PWAM$) was employed for the analysis. Not only the whole study area, but also each of its 318 prefecture-level cities were analyzed. We found that: 1) a considerable decrease of the

$PWAM$ has been observed during 2014-2020. 2) In 2020, the $PWAM$ is lower than 35 μg/m$^3$, not only for the whole study area but also for 214 of its prefecture-level cities, which accounts for 67% of its total population. The results indicates a considerable improvement of air quality in Eastern China. More important, this study illustrates the feasibility of using open geospatial data to monitor the SDG indicator 11.6.2, which can be applied to other countries and regions.

**Notes**

1. https://zenodo.org/record/4660858#.YQevavmt9zk

2. https://www.worldpop.org/

3. https://www.webmap.cn/main.do?method=index

4. https://sedac.ciesin.columbia.edu/data/set/sdei-global-annual-gwr-pm2-5-modis-misr-seawifs-aod/data-download

5. https://landscan.ornl.gov/

6. https://ghsl.jrc.ec.europa.eu/ucdb2018Overview.php